\documentclass{eptcs}
\usepackage{graphicx}
\usepackage{comment}
\usepackage{float}

\usepackage{graphicx}
\usepackage{amssymb}
\usepackage{epstopdf}
\newcommand{\ep}{\epsilon}
\newcommand{\bM}{\bf M}
\newcommand{\bK}{\bf K}

\newtheorem{definition}{Definition}

\floatstyle{boxed}
\newfloat{program}{thp}{lop}
\floatname{program}{Program}

\title{A short note on Simulation and Abstraction}

\author{Chris Hankin
\institute{Institute for Security Science and Technology\\
Imperial College London, UK}
\email{c.hankin@imperial.ac.uk}}

\def\titlerunning{Simulation and Abstraction}
\def\authorrunning{Chris Hankin}

\begin{document}
\maketitle

\begin{abstract}
This short note is written in celebration of David Schmidt's sixtieth birthday.  He has now been
active in the program analysis research community for over thirty years and we have enjoyed
many interactions with him.  His work on characterising simulations between Kripke structures using Galois connections was particularly influential in our own work on using probabilistic abstract 
interpretation to study Larsen and Skou's notion of probabilistic bisimulation.  We briefly review this work and discuss some recent applications of these ideas in a variety of different application areas.
\end{abstract}

\section{ Introduction}
Since his earliest contributions on state transition machines for lambda calculus expressions 
\cite{Schmidt80}, David Schmidt has been at the forefront of research in programming language
theory, particularly program analysis. 

His work on program analysis started through his collaboration with Neil Jones.
No doubt partly inspired by Patrick and Radhia Cousot's work on abstract interpretation \cite{CC77},
he has made a study of various aspects of Galois connections.  An early contribution
was \cite{MeltonSS85}.  A later example, which was influential in our own work, was 
\cite{Bisim99} where he shows how to characterise simulation relations using Galois
Connections.

The early work of our group was also inspired by the Cousots \cite{BHA86}.  Over the last 
fourteen years, we have been working on the analysis of probabilistic and quantitative
programming languages and systems \cite{DPW00}.  This has led to a framework called 
{\it probabilistic abstract interpretation} (PAI).  Rather than using lattices and Galois Connections,
PAI uses Hilbert Spaces and Moore-Penrose Pseudo Inverses.  Inspired by \cite{Bisim99},
we have used PAI to characterise probabilistic bisimulation \cite{DPHW03,DPHW05}; 
we demonstrated the probabilistc analogue of Dave's earlier results which amounted to
characterising Larsen and Skou's probabilistic bisimulation \cite{LS89} using Moore-Penrose
Pseudo Inverses.  A major feature of our approach is that it becomes natural to introduce 
a notion of approximate bisimulation -- this has proved to be very useful in studies of
language-based security \cite{DPHW04}.


The author first met David Schmidt nearly thirty years ago.  He visited Imperial College
whilst developing the work in \cite{MeltonSS85}, his visit coincided with our own work on
the ideas in  \cite{BHA86}.
We found a lot to talk about and it was the start of many further interactions.  In addition to his
scientific contributions, he has always given time to developing text books;  his \cite{Den86}
was used to educate several generations of Imperial students in the principles of 
programming language design.  The author has also worked with Dave on many
programme committees; perhaps the high point was when they were co-General Chairs
of POPL in 2001.  Dave has a had a long and successful career and we wish him many more years. 
The author would like to add his congratulations to Dave on this important milestone.

\section{Simulation and Galois Connections}

In \cite{Bisim99}, Dave addresses, {\em inter alia}, the question of characterising
simulation relations between Kripke structures using Galois connections.

Recall that  
we define $(L,\alpha,\gamma,M)$ to be a {\em Galois connection}
between the complete lattices $(L,\sqsubseteq)$ and $(M,\sqsubseteq)$
if and only if
\begin{center}
$\alpha: L \rightarrow M$ and
$\gamma: M \rightarrow L$ are monotone functions
\end{center}
that satisfy:
\[
\begin{array}{rcl}
\gamma \circ \alpha & \sqsupseteq & id_L
\\[1ex]
\alpha \circ \gamma & \sqsubseteq & id_M
\end{array}
\]

For Kripke structures $C = \langle \Sigma_C, \rightarrow_C, {\cal I}_C \rangle$
and $A = \langle \Sigma_A, \rightarrow_A, {\cal I}_A \rangle$, a binary relation
${\cal R} \subseteq \Sigma_C \times \Sigma_A$ is a {\em simulation} of {\it C}
by {\it A} ($C \lhd_{\cal R} A$), if for every $c \in \Sigma_C$, $a \in \Sigma_A$:
\[
\mbox{if}~c\ {\cal R}\ a\ \mbox{and}\ c \rightarrow_C c'\ \mbox{then}\ \exists a' \in \Sigma_A[
a \rightarrow_A a'\ \mbox{and}\ c'\ {\cal R}\ a']
\]

In the framework studied in \cite{Bisim99}, the concrete Kripke structures are often infinite
state structures representing programs and the abstract Kripke structures are some program
analysis.  In this setting, we should abstract sets of states of the concrete structure to a single
state in the abstract structure.  Given a Galois connection, $(\alpha : {\cal P}(\Sigma_C)
\rightarrow \Sigma_A, \gamma : \Sigma_A \rightarrow {\cal P}(\Sigma_C))$, we can construct the 
relation, ${\cal R}_{(\alpha,\gamma)} \subseteq {\cal P}(\Sigma_C) \times \Sigma_A$:

\[
S\ {\cal R}_{(\alpha,\gamma)} a\ \mbox{if and only if}\ \alpha(S) \sqsubseteq_A a
\]

This can be shown to be the suitable basis for a simulation relation.

Whilst \cite{Bisim99} achieves much more than described here, it was these ideas that
inspired our own work on probabilistic bisimulation (originally introduced by Larsen and Skou
\cite{LS89}) for reactive systems (also called fully probabilistic systems): 

\begin{definition}
  A {\em probabilistic bisimulation} is an equivalence relation $\sim_b$ on
  states of a probabilistic transition system satisfying for all actions
  $a \in A$:
\[
  p \sim_b q             \mbox{ and } p \rightarrow_a \pi
  \Leftrightarrow
  q \rightarrow_a \rho  \mbox{ and } \pi \sim_b \rho.
\]
where $\pi$ and $\rho$ are distributions of states.
\end{definition}.  

\section{Probabilistic Bisimulation and Moore-Penrose Pseudo Inverses}

In \cite{DPHW05,DPHW03}\ we introduced an approximate version of bisimulation
and confinement where the approximation can be used as a measure $\ep$ for the
information leakage of the system under analysis. 
We represented probabilistic transition systems by linear operators, i.e. by their transition matrices $\bM$. 
In the case of probabilistic programs and systems these matrices $\bM$ are the usual
well known stochastic matrices which are the generators of the corresponding
Markov chains (for details see \cite{DPHW05,DPHW03}).

We then showed that two systems $\bM_1$ and $\bM_2$ are bisimilar if there
exist simplified, or abstracted, versions of $\bM_1$ and $\bM_2$, represented by
matrices $\bM_1^\#$ and $\bM_2^\#$, such that $\bM_1^\# = \bM_2^\#$. 
In the probabilistic abstract interpretation setting that we use, a bounded linear operator and
its Moore-Penrose Pseudo Inverse  are the analogue of the adjoined pair of monotonic
functions in a Galois insertion.  The
abstract systems are obtained by {\em lumping} states, i.e. by identifying
each concrete state $s_i$ with a class $C_j$ of states which are all
behavioural equivalent to each other.

Concretely, we compute this via $n \times m$ matrices $\bK$ (where $n$ is
the number of concrete states and $m$ the number of abstract classes) with
$\bK_{ij} = 1$ iff $s_i\in C_j$ and $0$ otherwise. We refer to such matrices
which have exactly one entry $1$ in each row while all other entries are $0$
as {\em classification matrices}, and denote the set of all classification
matrices by $\cal K$. The abstract systems are then given by $\bM_i^\# =
\bK_i^\dagger\bM_i\bK_i$ with $\bK_i$ some classification matrix and $\dagger$
constructing the so called {\em Moore-Penrose pseudo-inverse} -- in the case
of classification matrices $\bK^\dagger$ can be constructed as the
row-normalised transpose of $\bK$.

The problem of showing that two systems $\bM_1$ and $\bM_2$ are behaviourally
equivalent, i.e. are (probabilistically) bisimilar, is now translated into
finding two classification matrices $\bK_i\in{\cal K}$ such that
\[
\bM^\#_1 =\bK_1^\dagger\bM_1\bK_1 = \bK_2^\dagger\bM_2\bK_2 = \bM^\#_2.
\]

In case that two systems are not bisimilar we can still define a quantity
$\ep$ which describes how (non-)bi\-si\-mi\-lar the two systems are. This
$\ep$ is formally defined in terms of the norm of a linear operator
representing the partition induced by the `minimal' bisimulation on the set of
the states of a given system, i.e. the one minimising the observational
difference between the system's components (see again \cite{DPHW05}\ for
further details, in particular regarding labeled probabilistic transition systems):
\begin{definition} 
  Let $\bM_1$ and $\bM_2$ be the matrix representations of two probabilistic
  transition systems. We say that $\bM_1$ and $\bM_2$ are {\em
    $\ep$-bisimilar}, denoted by $\bM_1 \sim_b^\ep \bM_2$, iff
  \[
  \inf_{\bK_1,\bK_2\in{\cal K}} 
  \| \bK_1^\dagger \bM_1 \bK_1 - \bK_2^\dagger \bM_2 \bK_2 \| = \ep
  \]
  where $\|.\|$ denotes an appropriate norm, e.g. the supremum norm
  $\|.\|_\infty$.
\end{definition}

In \cite{DPHW05} we show that, when $\ep =0$ this gives the standard notion of
probabilistic bisimulation.

\section{Conclusion}

This short note has sketched some early work by David which forms part of his deep study
of the use of Galois connections and relations in program analysis.  Our own work on
characterising probabilistic bisimulation using probabilistic abstract interpretation has found a number
of applications, including:
\begin{itemize}
\item the detection and removal of timing channels in probabilistic transition systems \cite{DPHW12}  --
we study a concept called probabilistic time bisimilarity and use it to detect timing channels;
\item the detection of sub-communities in social media \cite{LMH13} -- we evaluate a number
of algorithms including one
using the notion of stability from \cite{DYB10} which effectively lumps nodes together if their mutual interactions are ``stronger" than interactions outside the group; and
\item the abstraction of stochastic and Bayesian games to provide decision support in
cyber security \cite{HM13} -- where we hope to apply probabilistic abstract interpretation directly
to the underlying probabilistic transition systems in the games, thereby developing a principled way of reducing the state spaces to achieve tractability of game solutions.
\end{itemize}
We look forward to discussing some of this work with David in the future but, in the meantime, reiterate
our best wishes on this important anniversary.

\section{Acknowledgements}
Much of the work discussed above was done in collaboration with Alessandra Di Pierro and 
Herbert Wiklicky.  More recently, I have enjoyed working on the application of these ideas to other areas with
Erwan Le Martelot and Pasquale Malacaria.

\bibliographystyle{eptcs}
\bibliography{DPHW}

\end{document}